\documentclass[twocolumn,aps,showpacs,pra,groupedaddress]{revtex4}
\usepackage{epsfig,amssymb,amsmath}

\def\comment#1{}
\def\togli#1{}
\def\labell#1{\label{#1}}

\begin{document}
\vskip 1\baselineskip
\title{Architectures for a quantum random access memory}
\author{Vittorio Giovannetti$^1$, Seth Lloyd$^2$, Lorenzo Maccone$^3$}
\affiliation{$^{1}$NEST CNR-INFM \& Scuola Normale Superiore, Piazza
  dei Cavalieri 7, I-56126, Pisa, Italy.\\$^{2}$MIT, Research
  Laboratory of Electronics and Dept. of Mechanical Engineering,\\
  77 Massachusetts Avenue, Cambridge, MA 02139, USA.\\
  $^3$ QUIT - Quantum Information Theory Group, Dipartimento di Fisica
  ``A.  Volta'' Universit\`a di Pavia, via A.  Bassi 6, I-27100 Pavia,
  Italy.}  \date{\today}

\begin{abstract}
  A random access memory, or RAM, is a device that, when interrogated,
  returns the content of a memory location in a memory array. A
  quantum RAM, or qRAM, allows one to access {\it superpositions} of
  memory sites, which may contain either quantum or classical
  information.  RAMs and qRAMs with ${n}$-bit addresses can access
  $2^{n}$ memory sites.  Any design for a RAM or qRAM then requires
  $O(2^{n})$ two-bit {\em logic gates}.  At first sight this
  requirement might seem to make large scale quantum versions of such
  devices impractical, due to the difficulty of constructing and
  operating coherent devices with large numbers of quantum logic
  gates. Here we analyze two different RAM architectures (the
  conventional fanout and the ``bucket brigade'') and propose some
  proof-of-principle implementations which show that in principle only
  $O({n})$ two-qubit {\em physical interactions} need take place
  during each qRAM call.  That is, although a qRAM needs $O(2^{n})$
  quantum logic gates, only $O({n})$ need to be activated during a
  memory call. The resulting decrease in resources could give rise to
  the construction of large qRAMs that could operate without the need
  for extensive quantum error correction.
\end{abstract}
\maketitle 

Random access memory (RAM) is a highly versatile device for
storing and accessing information.
It consists of an array of memory cells  where
information is stored in the form of bits. 
Each cell is associated to
a unique address, i.e.~a number which identifies the location of the
cell.  The main characteristic of a RAM is that each memory cell
can be separately addressed at will, whence the designation `random access'.
To access a cell, its address must be provided in an input register
(the {\em  index register}). The device will then output the content
of the memory cell in a second register (the {\em output register})~\cite{ram}. 
If the input
register is composed by ${n}$ bits, the RAM is capable of addressing
$2^{n}$ different memory locations: 
when given an ${n}$-bit address $k$, the RAM
returns the bit string $f_k$ which was stored in the memory 
slot of the database labeled by $k$.

A quantum random access memory, or qRAM, is a RAM that functions in a
way that preserves quantum coherence~\cite{chuang}.  Compared to its
classical counterpart a qRAM has the additional feature that it allows
quantum superposition of addresses: given an input address state
$\sum_k \alpha_k |x\rangle_Q$ it returns the output state $\sum_k
\alpha_k |k\rangle_Q|f_k\rangle_A$ (here Q and A are, respectively,
the quantum analogous of the index register and of the output
register).  Just as classical RAMs are highly useful devices, quantum
random access memories will form an essential part of any large-scale
quantum computer.  Quantum random access memories that access {\it
  classical} information in quantum superposition could be used to
give exponential enhancements of a variety of data processing tasks,
such as pattern recognition \cite{pattern,pattern1,pattern2,pattern3}
and is necessary to implement quantum searching on a classical
database~\cite{chuang}.  More generally, a qRAM capable to access
either classical or quantum memory arrays is a fundamental ingredient
for many known and new algorithms, such as quantum
searching~\cite{grover}, collision finding~\cite{algorithm1},
element-distinctness in the classical~\cite{algorithm2} and
quantum~\cite{algorithm3} settings, the quantum algorithm for the
evaluation of general NAND trees~\cite{algorithm4}, or new algorithms
such as to enforce privacy in database searches~\cite{PQQ}, or to
route signals coherently through a quantum internet~\cite{qrouter}.

Conventional designs for classical and quantum RAMs require $O(N =
2^{n})$ two-bit or two-qubit logical operations in order to perform
one memory call.  At first sight this makes large scale quantum
implementations impractical without the use of extensive quantum error
correction. In this paper we show that there exists a routing
algorithm, the ``bucket-brigade'' introduced in Ref.~\cite{bb}, which
requires only a small number ($O({n})$) of two-qubit interaction to
implement a memory call.  Even though the number of logic gates in the
system is necessarily is exponential, only $O({n})$ of those gates
need be activated in the course of a qRAM memory call.  A classical
RAM that uses the bucket-brigade addressing schemes need only activate
$O(n)$ transistors in the course of a memory call, in contrast with a
conventional RAM that activates $O(2^n)$ transistors.  As a result, a
RAM that uses our design might operate with less dissipation and power
consumption than a conventional RAM. Note, however, that energy costs
in the memory addressing are not sufficiently high in current RAM
chips to justify an immediate adoption of the bucket-brigade. Other
sources of inefficiencies and dissipations are currently predominant
(mostly in the memory cells themselves). However, new promising memory
cell technologies are being developed (e.g.~the ``memristor''
cells~\cite{memristor}), which would drastically cut back cell
dissipation, so that cutting back dissipation in the addressing may
become important in the future. In the quantum regime, the addressing
scheme might allow the construction of a relatively large quantum RAM
without the need for costly quantum error correction.

The outline of the paper follows. We start by analyzing the two
different RAM architectures: the conventional (fanout)
architecture~\cite{chuang} and the bucket-brigade
architecture~\cite{bb}. In the latter, the control lines that route
the signals to the destination memory cell are replaced by memory
elements that control the routing.  The result is an exponential
decrease in the number of two-body interactions required to implement
the routing and to address memory.  Both these architectures can be
used in the quantum regime.  We conclude the paper by giving some
physical implementations of both qRAM schemes.

\section{Description of the protocol}\labell{s:proto}
To describe the qRAM algorithm, it is convenient to start from the
analysis of its classical counterparts.

In the conventional `fanout' addressing scheme, the index register
specifies the direction to follow to reach the memory cell we are
interested in. In particular, writing the index register in binary
form, each bit of such register can be interpreted as the direction to
take at a bifurcation of a binary tree. Since the ${n}$ bits encode
the directions to take at ${n}$ bifurcations, then $2^{n}$ different
paths can be described this way.

A realization of such indexing procedure is presented in
Fig.~\ref{f:ram1} part a).  We call it the `fanout' RAM scheme, since
each address bit fans out to control several switches placed on
different nodes of the tree.  In particular the $k$'th bit of the
index register acts as a controller for $2^k$ switches: if the bit has
value 0 {\em all} the switches under its control will point ``up'',
otherwise they will all point ``down''.  By looking at
Fig.~\ref{f:ram1}, it is easy to see that for every state of the index
register only one out of the $2^{n}$ possible paths has all the
switches in a ``conducting'' state: a signal will follow the path to
the memory cell we are interested in.  The main drawback of the fanout
architecture is that it requires simultaneous control over all the
$2^n-1$ nodes of the binary tree even though only ${n}$ nodes directly
participate to the addressing of a given memory cell (these are the
switches which lie along the path to the addressed memory cell).
Fanout schemes are commonly implemented in RAM chips~\cite{ram} by
translating them into electronic circuits where the switches of the
binary tree are replaced by pairs of transistors --- see
Fig.~\ref{f:ram1}b). Notice that the number of activated transistors
can be reduced to $O(n)$ with a more clever arrangement. In
Fig.~\ref{f:ramcl2} we modify the fanout RAM to attain a circuit where
at each level of the binary tree only two transistors are activated,
and where at the final stage only a single transistor is activated to
open up the unique path to the desired memory slot. In total only
$2n+1$ transistors are activated for every memory call. However, it is
important to note that the few last bits in the index register are
connected (in the last levels of the tree) to an exponential number of
transistors. Hence, this architecture would not be more helpful in the
implementation of a quantum RAM than the conventional fanout
architecture, as they both require the maintenance of quantum
coherence over an exponential number of connections.

 \begin{figure}[t]
\begin{center}
\epsfxsize=.8\hsize\leavevmode\epsffile{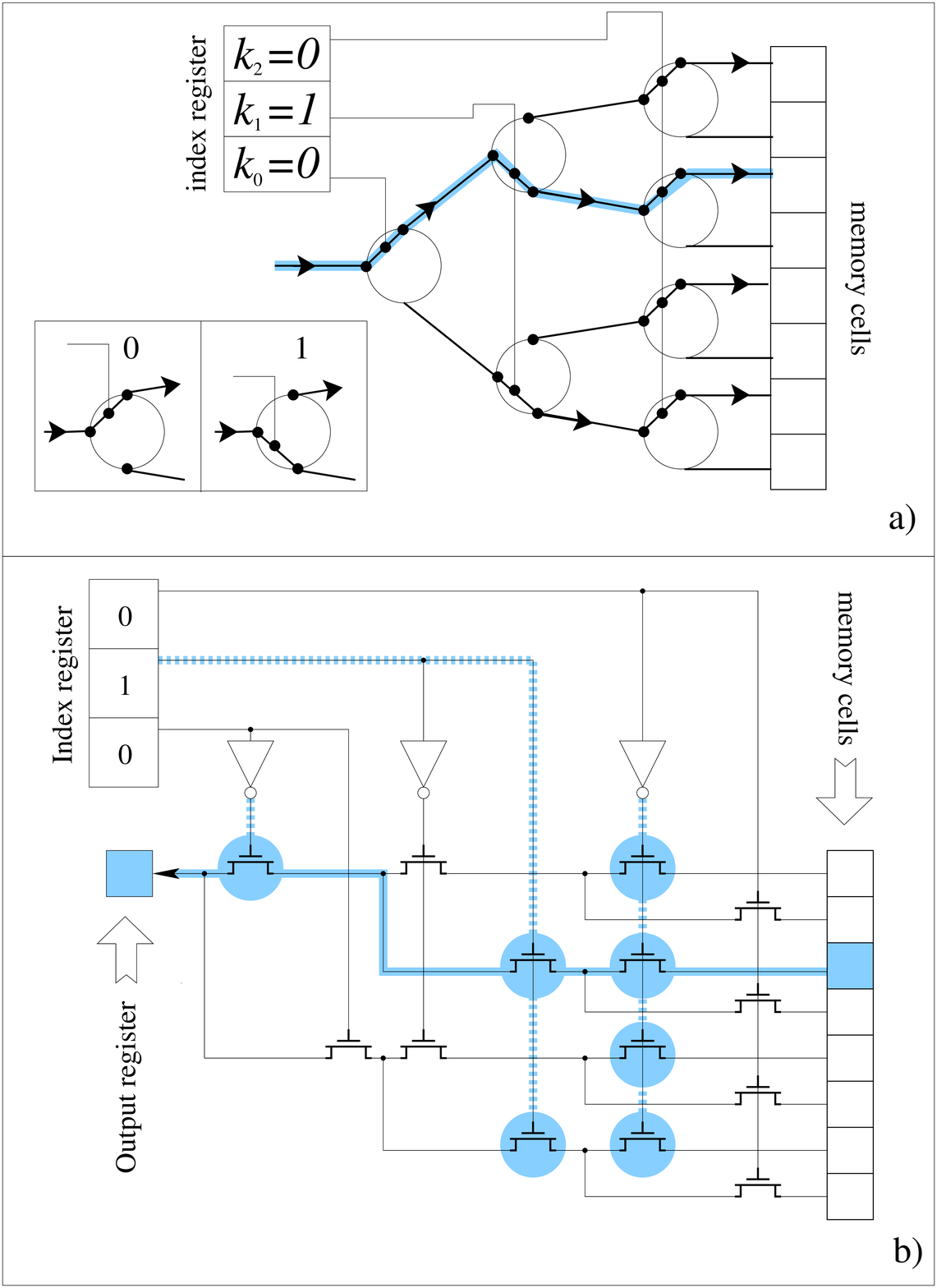} 
\end{center}
\caption{(Color online) Part a): Schematic of the fanout addressing
  scheme.  The bits of the index register control the switches (black
  circles in the picture) placed at the nodes of a binary tree.  The
  thick path corresponds to the address where all the switches are in
  a conducting state. It leads to the memory cell that is to be
  addressed.  Part b): typical electronic implementation~\cite{ram}.
  Each bit of the index register is encoded via dual rail logic in the
  conducting states of two electric wires (vertical lines in the
  picture).  The logical switches are represented by pairs of
  transistors that are controlled by the index register through the
  vertical wires (the total number of transistor is $2 (2^n-1)$).  The
  output register is connected to the memory cells through $2^n$ paths
  each containing $n$ transistors.  For each value of the index
  register only one of the paths has all transistors active (the thick
  path in this example).  For each memory call half of the transistors
  of the system are activated (the circled one in the picture) but
  only $n$ of them directly participate to the addressing (the ones
  which lie on the thick path).  The triangles represent NOT gates.  }
\labell{f:ram1}
\end{figure}
 \begin{figure}[t]
\begin{center}
\epsfxsize=.8\hsize\leavevmode\epsffile{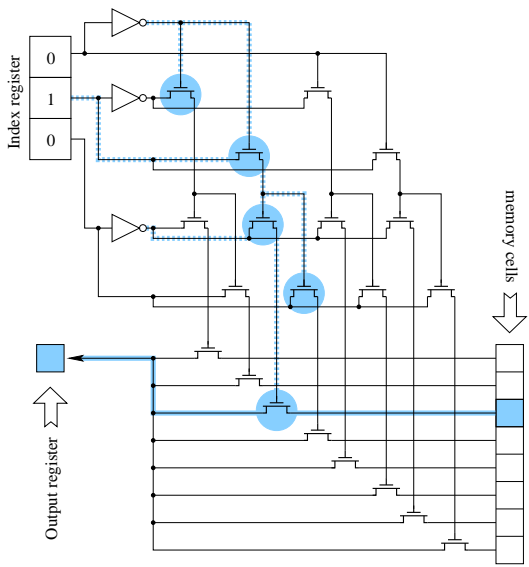} 
\end{center}
\caption{(Color online) Modification of the conventional fanout
  circuit of Fig.~\ref{f:ram1}b. This architecture achieves memory
  addressing with the activation of only $O(n)$ transistors. }
\labell{f:ramcl2}
\end{figure}

An alternative addressing scheme~\cite{bb} 
can be designed by changing the
function of the $2^{n}-1$ switching elements in the binary tree.  We
name this procedure ``bucket-brigade'' since both the routing and the
input-output signals are passed along the same route, like buckets of
water passed along a line of improvised fire-fighters. In each node of
the binary tree we place a three-state logic element (a ``trit''),
which can store the values $0$, $1$ or $\bullet$, where $\bullet$ is a
passive or `wait' state --- see Fig.~\ref{f:bb}a). A trit in the state
$0$ or $1$ acts as a switch that routes the signals transiting through
its node ``up'' or ``down'', respectively.  A trit in the state
$\bullet$ does not propagate the signals transiting through its node,
but changes its state to match the one of the arriving bit.  Initially
all the trits are in the `wait' state $\bullet$.  Now each bit of the
index register is sent into the binary tree, one at the time.  When
one of these bits encounters a trit in the state $\bullet$, its value
is transferred to the trit, which enters one of the two switching
states, $0$ or $1$.  A trit in the state $\bullet$ that receives a
$0$, itself becomes $0$, and routes all subsequent signals along the
$0$ branch of its node of the binary tree.  Similarly, a trit in the
state $\bullet$ that receives a $1$ becomes $1$, and routes all
subsequent signals along the $1$ branch.  After all the ${n}$ bits of
the index register have been sent through the binary tree, a route has
been carved through it: $2^{n}-({n}+1)$ of the trits in the binary
tree still have value $\bullet$, but any incoming signal will only
encounter the ${n}$ trits with value $0$ or $1$ which will route it to
the one destination (out of the $2^{n}$ possible ones) that was
initially encoded in the index register.
\begin{figure}[h]
\begin{center}
\epsfxsize=.8\hsize\leavevmode\epsffile{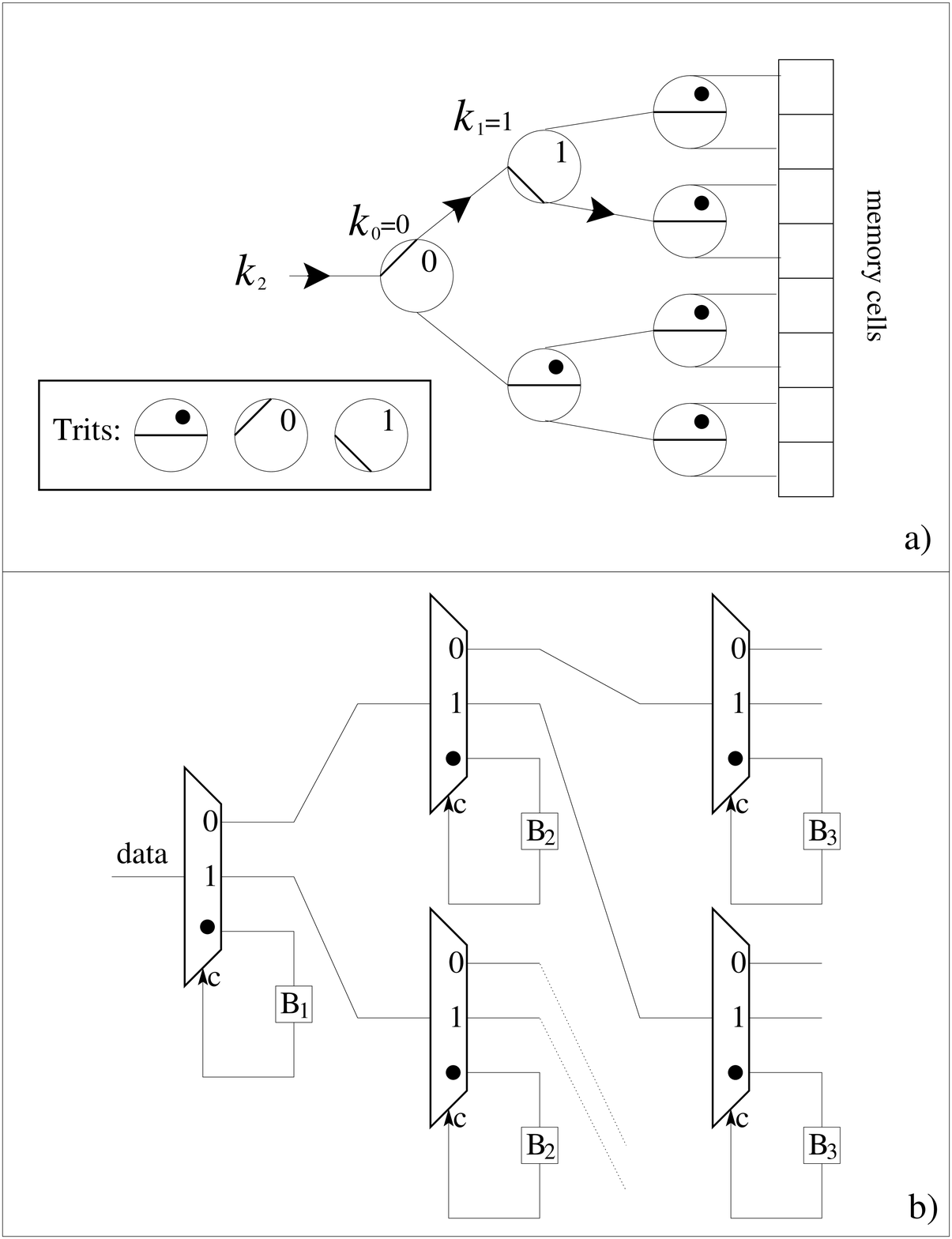}
\end{center}
\caption{Part a): Schematics for the bucket-brigade addressing scheme~\cite{bb}.
  Each node of the binary tree is a trit i.e. a logic element
  characterized by the internal states $0$, $1$ and $\bullet$. A trit
  in $0$ or $1$ acts as a switch that routes the signals transiting
  through its node ``up'' or ``down''.  A trit in the $\bullet$ state
  do not propagate the signals transiting through its node, but
  changes its state to a logical state $0$ or $1$, to match the one of
  the arriving bit.  Part b): Electronic implementation of the
  bucket-brigade. The trapezoids are multiplexers based on tri-state
  logic~\cite{art} which route any incoming signal to the output
  determined by the control lines ``c''. Suppose that initially all
  the memory elements B$_1$, B$_2$, etc. are in the state~$\bullet$.
  Any incoming signal is then routed to the memory element itself,
  which is then switched to the value of the incoming signal. Any
  subsequent signal will be routed along the 0 or 1 paths depending on
  the value that had arrived previously. An internal clock resets all
  the memory elements to the value $\bullet$ after the $2n$ clock
  cycles necessary for the memory addressing, so that the protocol can
  restart.}  \labell{f:bb}
\end{figure}

\subsection{Quantum RAM}
Both the fanout scheme and the bucket-brigade scheme
can be turned into quantum RAM algorithms by
transforming them into reversible processes and by requiring
quantum coherence to be preserved.  In abstract terms this
is done by sending a quantum signal (the {\em quantum bus})
back and forth in  the binary trees 
of Fig.~\ref{f:ram1} and \ref{f:bb},
and by coupling it with the memory cell through C-NOT transformations.  

We will see that in the quantum version of the fanout RAM, the $k$'th
index qubit is still required to control $2^k$ bifurcations in the
binary switching tree, creating a fragile macroscopic superposition.
While such high fanout control gates are relatively straightforward in
a classical setting, they are rather costly in a quantum setting,
where amplifying a signal makes it prone to decoherence.  In
particular, a qubit that interacts with $2^k$ quantum switches is
highly likely to decohere as $k$ increases, even if its interaction
with any given switch is small.  In Sec.~\ref{s:optical} we will
present optical and solid state implementations of high fanout control
gates.  Because of their exponential character such gates must
inevitably succumb to problems of noise and decoherence as ${n}$ gets
large.  The gates of the fanout architecture must then introduce
errors which scale as $O(2^{-n})$ (as the first gates in the
bifurcation tree are responsible for an exponential number of paths,
while the last gates in the bifurcation tree are coupled to an
exponential number of paths). Nonetheless for small ${n}$, useful
fanout qRAMs might be implementable along the lines proposed.  In
addition, fanout qRAM designs are instructive as they allow us to
identify failure mechanisms as ${n}$ gets large.

In contrast, the quantum version of the bucket-brigade addressing
scheme does not suffer from this problem: even though the number of
components of the device is $O(2^n)$, the number of control gates that
needs to be performed in each memory call is only polynomial in $n$.
As a result, bucket-brigade qRAMs might be scalable to large sizes,
even with relatively high switching error rates.  If the error rate
per switching event is $\epsilon$, then the overall error rate per
memory call is ${n}\epsilon = \log_2 N \epsilon$. This implies that
the bucket-brigade can operate also if its gates introduce errors
which scale as $O(1/n)$, in contrast to the fanout architecture. For
example, a per-switch error rate of $1\%$ allows one to address
$2^{10} \approx 10^3$ memory locations in quantum superposition with
an overall error rate of $10\%$, $2^{20} \approx 10^{6}$ memory
locations with an overall error rate of $20\%$, and $2^{30} \approx
10^{9}$ memory locations an overall error rate of $30\%$.  Because of
the exponential improvement of the bucket-brigade over fanout designs,
a small change in the error rate can lead to a dramatic improvement in
performance.  A per-switch error rate of $0.1\%$ allows one to address
$2^{100} \approx 10^{30}$ memory locations in quantum superposition
with an error rate of $10\%$.

In addition to employing the fanout or the bucket-brigade architectures,
it is possible to use a combination of the two.
In this case, there is a trade-off between
the amount of coherent control and the number of quantum memory
elements that are necessary.

\subsubsection{Fanout quantum RAM}
To quantize the fanout RAM, we employ a quantum bus. This is 
a quantum system that can be routed coherently along every one
of the $2^{n}$ possible paths of the binary tree and which
is characterized by some internal degree of freedom 
that can be used to store and process information.
At each bifurcation in the path,
the choice of the route (``up'' or ``down'') is performed in a
quantum-controlled fashion, i.e.~by using a unitary routing transformation
controlled by the corresponding qubit in the index register.  As
in the conventional fanout RAM, the $k$'th index qubit in index register 
of a fanout qRAM controls $2^k$ bifurcations.   

The ${n}$ quantum controlled transformations along the binary tree
convert the binary value of the index register Q into the position of
the bus system.  This transformation is a binary-to-unary translation:
the binary address stored in the index register is translated into the
unary variable consisting of the position of the bus.  The bus system
then locally interacts with the memory cell placed at the exit of the
path that it followed.  If the index register Q contains a
superposition of addresses, then the bus system follows more than one
path in superposition and locally interacts with all the memory cells
relative to these paths.  (More precisely, the position of the bus
becomes entangled with the state of the index register.)  This
interaction coherently copies the content of the memory cell into the
internal degree of freedom of the bus system. In the case in which the
each memory cell of the database contains a single bit of information,
the quantum bus is a single flying qubit and the copy operation is a
single C-NOT gate. If, instead, each memory cell contains $d$ bits of
information we can either use a single quantum bus with $2^d$ internal
states, or we can transfer the information bit by bit with a single
two-dimensional quantum bus by repeating the whole process $d$ times.

At this point, even though the desired information has been
transferred to the bus, the user cannot yet access it in quantum
superposition, since the value of the Q register is
still correlated (more precisely, entangled) 
with the position of the bus system.  Since we want to
preserve the quantum coherence, this correlation must first be
removed.  This can be achieved by ``uncomputing'' the binary-to-unary
translation, i.e.~by running the translation backwards.  In practice,
this means that the bus system must be sent back along the
quantum-controlled binary tree: at each bifurcation the controlled-Unitary
transformations decorrelate the value of the controlling qubit in the
${n}$ register from the bus system, which is thus routed back to the
binary tree entrance.  Now, we just swap the state of the bus's internal
degree of freedom to the output register A.  The final result is that
the uncomputation has decorrelated the bus system from the index
register Q, and the answer register A contains the state of the memory
cell (cells) addressed by Q.  If Q is initially in a superposition
of address states, after the quantum memory call has been completed,
Q and A are in an entangled state, with each address in Q perfectly
correlated with the corresponding output in A.

\subsubsection{Bucket-brigade quantum RAM}

To quantize the bucket-brigade procedure, in each node of the
binary tree we place a three-level quantum system (a ``qutrit''). If
it is in the state $|0\rangle$ or $|1\rangle$, then the qutrit acts as a
quantum switch, routing the subsequent incoming qubits ``up'' or
``down'', respectively.  If, instead, the qutrit is in the state 
$|\bullet\rangle$,
then a unitary operation $U_s$ stores any incoming qubit by swapping
the qubit's state into the state of the qutrit's first two levels. The
qubits of the Q register are sequentially sent through the apparatus:
the state of the $k$th index qubit is stored in exactly one qutrit at 
the $k$th level
of the tree and routes the subsequent qubits according to its state.
After all the ${n}$ qubits of the Q register have gone through the tree,
${n}$ quantum switches are active (i.e. in a state $|0\rangle$ or
$|1\rangle$).

At this point, we send a bus qubit
into the tree: it reaches the end of the tree where it meets with one
of the $2^{n}$ memory cells (when the index register in is a superposition
of addresses, the bus qubit can meet more than one cell, in superposition).
Now the state of this memory cell is coherently copied into the bus. 
The uncomputation needed to
decorrelate the position of the bus qubit can be easily performed by
reflecting it back into the binary tree: the qutrit-switches route it
back to the entrance of the tree.  Finally, the loading of the qutrits
must be reversed.  Starting from the last level of the tree, the
unitary ${U_s}^\dag$ is applied so that the qubits of the Q register
are recovered one by one and routed back to the entrance of the tree.
After ${U_s}^\dag$ has been applied to all levels of the binary tree,
the Q register has been recovered and the qutrits in the tree have all
been restored to the state $|\bullet\rangle$.  At the conclusion of the
entire process, the Q register
is decorrelated from the binary tree and the bus qubit contains the
value of the accessed memory cells.  Notice that, even
though this procedure requires the existence of a total number of qutrits 
of the order of $2^{n}$, only ${n}$ of them are active in any 
run of the protocol.

\subsection{Generalizations}\labell{sec:gen}

It is worth pointing out that the qRAM schemes we have presented in
the previous sections do not require the information contained in the
addressed memory cells to be classical. Indeed the whole architecture
still works also in the case in which the information stored in the
memory cells is purely quantum.  Of course in this case the no-cloning
theorem~\cite{WZ} forbids us to ``copy'' the quantum data in the
output register A: the protocol will simply delocalize the information
originally contained in the addressed memory cell into an entangled
configuration of the cells and A. In fact, the procedure that copies
classical information, i.e.~the C-NOT gate, in general will entangle
quantum information on which it acts. Alternatively, a swap operation
can be performed to transfer the quantum information to the bus.  The
bus can then transfer the quantum information out of the qRAM, at
which point it can be sent to some desired destination. The type of
quantum data processing that one implements on the memory states of
the qRAM depends on the application one is interested in. Notice that,
in contrast to the case of classical memory cells, a qRAM procedure
based on either the swap or the C-NOT transformations will generally
leave the memory array entangled with the index and output registers 
Q and A at the end of the procedure.

Before proceeding we note also that it is possible to reduce the
number of circuit elements in all the above implementations by using
more clever geometric arrangements of the memory array. Typically, a
bidimensional array is used, and the index register is divided into
two parts, one of which indexes a row in the memory array, the other a
column. If rows and columns have the same size, we can reduce the
number of necessary circuit elements to the square root of its
original value, and still address the same number of memory cells.
More complex arrangements (such as using three-dimensional memory
arrays) would entail a further reduction of circuit elements, but are
not typically used as they require much more complex wiring diagrams.
With this in mind, in the rest of the paper we will concentrate on the
case of one-dimensional memory arrays for the sake of simplicity.

\section{Physical implementations}\labell{s:optical}

The basic concept of a qRAM dates to the early days of quantum
computing, when researchers first began considering quantum
computer architectures: a sketch of an optical fanout qRAM scheme 
can be found in Ref.~\cite{chuang}.
In this section we construct detailed physical designs for
qRAMs, taking into account the possibility of hybrid encodings where 
the quantum bus and the memory cells are implemented by different 
physical objects (e.g. photons vs. trapped atoms).
Finally, we present an implementation of the bucket-brigade scheme.

\subsection{Quantum optical implementation of the fanout qRAM} 
\labell{sec:quantumopt}
An optical implementation of the fanout qRAM is presented in 
Fig.~\ref{f:polarizz}. For the sake of simplicity we consider
the case in which the memory cells contain a single bit.
\begin{figure}[t]
\begin{center}
\epsfxsize=.8\hsize\leavevmode\epsffile{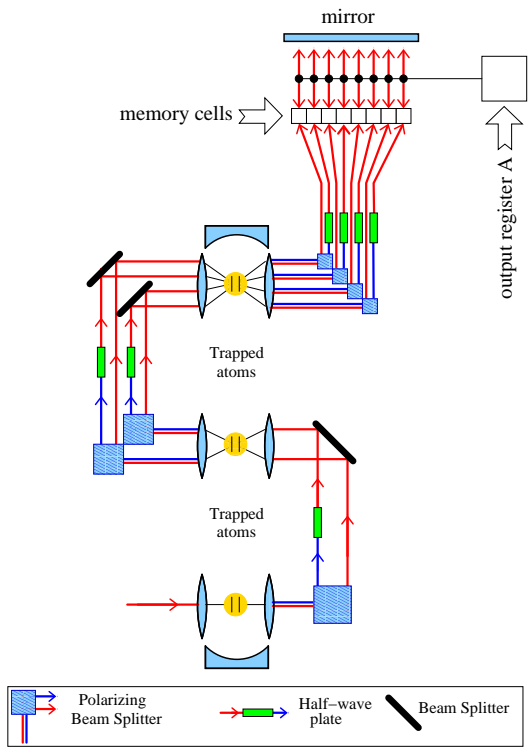}
\end{center}
\vspace{-.5cm}
\caption{(Color online) Schematic of the optical implementation of
  fanout qRAM model.  The index register is encoded into the internal
  state of a collection of $n$ two-level atoms trapped (the gray
  circles in the picture) in an optical lattice.  The quantum bus is a
  single photon which can be prepared in two orthogonal polarizations
  (represented in the pictures by the double lines) and routed into
  $2^n$ spatial modes. The routing algorithm begins by preparing the
  bus photon in one polarization state and by coupling it with the
  first atom of the trap. This copies in a coherent fashion the
  internal state of the atom in the polarization on the photonic bus.
  Using a polarizing beam splitter and a half-wave plate this
  information is then transfered into a spatial degree of freedom. The
  resulting two modes are then coupled to the next atom of the index
  register and the whole process is repeated $n$ times. At the end of
  the routing, the atomic state is coherently mapped into the spatial
  degree of freedom of the bus photon. The photon then reaches the
  memory cells, where their information content is copied onto its
  polarization by a C-NOT operation (which does not produce
  entanglement, as the memory cells are in a classical state, 0 or 1).
  Then, its polarization state is swapped with the output register A
  which thus contains the value (or values, in superposition) of the
  addressed memory cells.  Finally, the photon (its polarization
  having been reset by the swap) is reflected back through the setup:
  the interactions with the atoms reverse the routing algorithm so
  that the photon in a superposition of modes is routed back to the
  single mode from where it had originally entered the setup.}
\labell{f:polarizz}\end{figure}
Here the value of index register Q
is encoded in a binary fashion in the internal state of $n$ atoms
(ions) trapped into an optical lattice (magneto-optical trap).
The $2^{n}$ memory locations are addressed  
by translating this information 
on $2^{n}$ different spatial modes of the electromagnetic field
of a `bus' photon characterized by polarizations $|0\rangle$ and $|1\rangle$.

We start with 
 the register Q
  prepared into the internal state of the ${n}$ atoms of the following form:
  \begin{eqnarray} |\Psi\rangle_Q= \sum_{k=0}^{2^{n}-1}\alpha_k
    |k_0\rangle_{Q_0}|k_1\rangle_{Q_1}\cdots|k_{n-1}\rangle_{Q_{n-1}}\labell{enc}\;,
\end{eqnarray}
where $\alpha_k$ is the amplitude that the register Q indicates the
$k$'th memory cell, and where $k_i$ denote the bits of $k$,
i.e.~$k=k_0k_1\cdots k_{{n}-1}$.  We prepare a single photon in the
polarized state $|0 \rangle$ and shine it onto the first atom: The
coupling among them is such that the atom acts as a controller for the
polarization of the photon (see Fig.~\ref{f:cnot}).  Then we send the
bus photon through a polarizing beam splitter and rotate its
polarization back to $|0\rangle$ (see Fig.~\ref{f:polarizz}).  As a
result, the spatial mode occupied by the bus photon is now perfectly
correlated with the value of the first index qubit, through a C-NOT
transformation.  We repeat the above procedure, using the second atom,
i.e.~apply a C-NOT gate using the qubit $|k_1\rangle$ as the control
and the bus photon as the target.  Note that the C-NOT transformation
must be crafted in such a way as to be independent of the position of
the target photon, as the bus photon is now in two different spatial
modes.  This can be achieved by letting the two spatial modes interact
with the gate at different times. Now again we can split the bus
photon's spatial mode using polarizing beam splitters and then reset
its polarization using polarization rotators, as before. The value of
the first two qubits $k_0$ and $k_1$ has been transferred to four
spatial modes (see Fig.~\ref{f:polarizz}).  We then iterate the
procedure until the values of all the ${n}$ qubits of index register
have been transferred to $2^{n}$ spatial modes of the bus photon.
Each of the $2^{n}$ spatial modes of the bus photon interacts with the
first bit of the corresponding location in the memory array. The
interaction is such that the bus photon's polarization is rotated if
the memory bit is 1. Now we swap the state of the bus photon's
polarization into the output register A, which was previously
initialized in $|0\rangle$.  Since the polarization of the bus photon
has been reset to $|0\rangle$, to invert the binary-to-unary encoding,
it is sufficient to reflect the photon back through the whole
setup~\cite{nota}.

\begin{figure}[t]
\begin{center}
  \epsfxsize=.5\hsize\leavevmode\epsffile{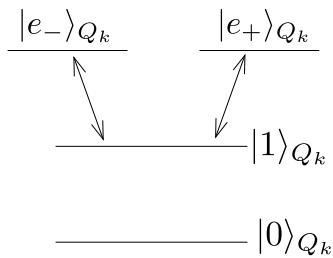}
\end{center}
\caption{Atomic level structure of the memory index elements. 
The $k$'th qubit of the index register  is stored in the stable levels
$|0\rangle_{Q_k}$ and  $|1\rangle_{Q_k}$ of the $k$'th atom. 
The atomic state $|0\rangle_{Q_k}$ does not interact with the
  photon. On the contrary, the state $|1\rangle_{Q_k}$, 
interacts  with a
  $|\pm\rangle=(|0\rangle \pm |1\rangle)/\sqrt{2}$-polarized photon 
through  a standard  
Jaynes-Cumming Hamiltonian which couples it with
the ancillary levels $|e_{\pm}\rangle_{Q_k}$. The 
interaction time is such that  the system 
 experiences a cyclic transition 
$|1\rangle_{Q_{k}}|\pm\rangle \rightarrow 
|e_\pm\rangle_{Q_k}|\mbox{vac} \rangle 
\rightarrow \pm |1\rangle_{Q_k}|\pm\rangle$~\cite{kimble} (here
$|\mbox{vac}\rangle$ is the photonic vacuum).
Consequently, a photon with polarization $|0\rangle$ 
will get entangled with the atomic levels $|0\rangle_{Q_k}$ and 
$|1\rangle_{Q_k}$.} 
\labell{f:cnot}
\end{figure}

Notice  that, even though the $k$'th address bit fans out to
$2^k$ controlled-controlled NOT gates, it only interacts with the
bus qubit once: the number of physical interactions between
information carrying degrees of freedom is exponentially smaller
than the number of physical gates.   This exponential-small paucity of 
interactions mirrors the classical case: in a classical fanout RAM,
even though the $k$'th address bit interacts with two sets of $2^k$
transistors, the actual signal sent through the RAM to address the
memory interacts with only ${n}$ transistors.    

\togli{The relatively small number of interactions between the bus
  degree of freedom and the address degrees of freedom implies that,
  although the number of necessary gates scales exponentially with the
  number ${n}$ of qubits in the query, the error deriving from the
  application of these gates can scale only linearly in ${n}$,
  depending on the form of the interactions between bus and address
  qubits.  If this interaction is `passive', so that errors occur only
  when an address qubit interacts with a bus photon, then the error
  rate scales as ${n}$.  If, by contrast, the error source is active
  (e.g., if each C-NOT introduces photons in the system, even if the
  mode on which it is performed is the vacuum), then the scaling will
  be different.  In particular, if each interaction of an address
  qubit with a vacuum mode of the bus photon induces even a small
  level of decoherence, as is typically the case, then the error rate
  grows exponentially in ${n}$.  A much better error resilience is
  given by the bucket-brigade procedure.}

\subsection{Phase gate implementation of the fanout qRAM}\labell{s:ssqram}

In this section we present a fanout qRAM  implementation
where the addressing is realized by a collection of 
controlled phase gates. 

The main idea  is to
implement the binary tree  as $2^n$ cavities in which (at most) one
photon is present (see Fig.~\ref{f:ssqram}).  The ${n}$ qubits of the Q register fan
out to induce phase shifts in all the cavities.  If these shifts are
chosen appropriately, the incoming photon can resonate in only the cavity
indexed by the register Q.  This can be achieved by using conditional
phase shifters (the solid squares of the picture)
 controlled by the index register Q. 
Since there are $2^{n}$ possible cavities,
we need to devise a scheme which is extremely sensitive to phase
shifts: the width of the cavity resonance window must be of the order
of $\pi/2^{n}$, so that only one among cavities with similar phase
shifts can be placed into resonance~\cite{NOTAdue}.  
Such routing scheme can be realized in principle both in
an  optical environment analogous to the one of Sec.~\ref{sec:quantumopt}
or in a solid state
implementations~\cite{wallraff1,wallraff2}.  In the latter case
 photolithographic techniques permit the etching
of micrometer scale solid-state qubits, e.g., superconducting charge
or flux qubits, or electron spin quantum dots.  These qubits can in
turn be coupled to microwave cavities~\cite{wallraff1,wallraff2}.
While existing systems couple only a few qubits to the cavity, because
of the small scale of the qubits (ranging from tens of nanometers for
superconducting charge qubits or quantum dots to tens of micrometers 
for superconducting flux qubits) compared with the
relatively large scale of the cavity (centimeters), in principle
microwave cavities can be constructed so as to interact with a large
number of such qubits simultaneously.  The interaction of a single
microwave photon with a large number of qubits can then be used to
implement the large scale fanout required for a fanout qRAM. 
In this case the conditional phase shifters 
can be implemented, for example, following Ref.~\cite{wallraff2} 
and by requiring a single superconducting qubit
which interacts with two microwave modes (one encoding the quantum
bus and one encoding the index register degree of freedom). 
In this case the required effective Hamiltonian could be 
of the Jaynes-Cummings form.  We need to operate the system far
from resonance to ensure that the microwaves modes do not exchange
excitations through this coupling ---  its only effect should
be the insertion of a phase shift conditioned on the value of the qubit.
\begin{figure}[hbt]
\begin{center}
  \epsfxsize=.9\hsize\leavevmode\epsffile{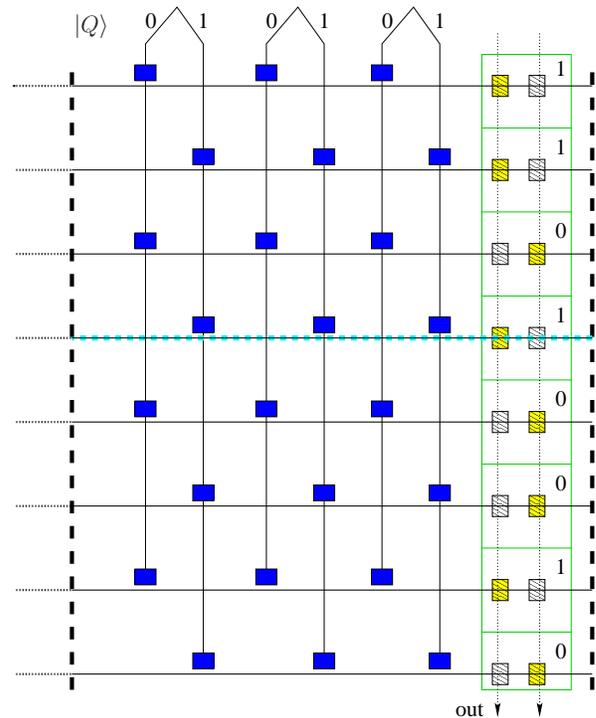}
\end{center}
\vspace{-.5cm}
\caption{(Color online) Phase gate implementation of the qRAM
  protocol. The dashed lines at the right and at the left indicate the
  microwave cavity. The solid rectangles indicate phase shifters
  controlled by the photons the Q register is encoded in (above). The
  hatched rectangles indicate phase shifters controlled by the cavity
  photon.  These phase shifters are implemented by a qubit (see text),
  the state of which is determined by the memory cell contents. In
  this example the Q register is in the state $|Q\rangle=|011\rangle$,
  so that the system is addressing the fourth memory location (which
  contains a one): the only path in resonance is the one indicated by
  the thick dashed line.}  \labell{f:ssqram}
\end{figure}

Once the routing has been performed the information is extracted from
the memory cells by coupling each cell to one of the $2^n$ cavities.
In the implementation discussed above, for instance, this can be
achieved by placing in each of the memory cell locations two
superconducting qubits that are employed both to store the information
in the memory cell and to extract this information to an outgoing
photon.  (Two qubits are needed in each cavity instead of one, in
order to remove the feedback effect that the content of the memory
cell would have on the cavity photon).  If the cell stores a ``one'',
then the first superconducting qubit in the state $|1\rangle$ and the
second is in $|0\rangle$. This means that the first qubit induces a
phase shift on the first outgoing photon, whereas the second qubit
leaves the second outgoing photon untouched.  If, instead, the cell
stores a ``zero'', then the states of the two superconducting qubits
are reversed.  With this procedure, the contents of the memory cell
have been transferred to the two outgoing photons: the first is
phase-shifted only if the memory contained a ``one'', while the second
is phase-shifted only if it contained a ``zero''.

The circuit model of Fig.~\ref{f:ssqram} can be used to construct
a variety of designs for solid-state fanout qRAMs.  A qRAM that
is effectively the direct quantization of the classical fanout
RAM could be constructed using single electron transistors
and Coulomb blockade.  For example,
the index qubits could correspond to single electron transistors 
which, when activated, allow charge coherently to tunnel onto charge
qubits, corresponding to the solid rectangles. 
These rectangles, when charged, in turn exert a Coulomb blockade
on their corresponding lines in the binary switching tree. 
The quantum bus is implemented by a single electron, which
is routed via the Coulomb blockade along the single unblocked branch 
of the tree.
Of course, the difficulties in performing all the Coulomb blockade
and tunneling steps in a coherent manner are formidable.  
Because the logical qubits in this system are all registered
by the presence or absence of electrons, charge noise is likely
to decohere the charge fanout qRAM very rapidly.  In the 
cavity solid state qRAM described in detail above, by contrast,
coherence times are set by the coherence scale of the optical
cavities used to perform the fanout and addressing.  Such coherence
times can be relatively long compared with charge qubit coherence times
(microseconds compared with nanoseconds).

\subsection{Bucket-brigade
  implementation}\labell{s:brigade} 
In this Section we give a possible implementation of the quantum
bucket-brigade scheme where the index register qubits are encoded into
photons propagating along a network of coupled cavities that contain
trapped atoms.  The main idea is the following.  In each node of the
binary tree, the qutrit is implemented by an atom with the level
structure given in the inset of Fig.~\ref{f:bb11}. The levels
$|zero\rangle$ and $|up\rangle$ are coupled to the ``up'' spatial
paths in the tree (i.e.~the dashed lines in Fig.~\ref{f:bb11}), while
the levels $|one\rangle$ and $|down\rangle$ are coupled to the
``down'' spatial paths (i.e.~the dotted lines).  Start the protocol by
preparing all atoms in the level $|\bullet\rangle$ and sending in a
photon which encodes into its polarization state the first qubit of
the Q register (alternatively one can replace polarization encoding
with time-bin encoding). By turning on a strong laser field to induce
a Raman transition, the photon will be absorbed and stored in the
first atom in the $|zero\rangle$ atomic level if it is in the state
$|0\rangle$, and in the $|one\rangle$ level if it is in $|1\rangle$.
Now send the second photon which encodes the second qubit of Q: it
will meet the atom in the state $|zero\rangle$ or $|one\rangle$
depending on the value of the first qubit.  Again we use Raman
transition technique to absorb (and subsequently re-emit) the photon
in the $|up\rangle$ level or in the $|down\rangle$ level respectively.
In the first case (since $|up\rangle$ and $|zero\rangle$ couple only
to the ``up'' spatial path that connects to the subsequent level in
the tree), the second qubit will move in the ``up'' path to the
uppermost atom in the second node.  Here it will be absorbed by the
$|zero\rangle$ or $|one\rangle$ level of the second atom, depending on
its value.  In the second case, instead, the qubit follows an
identical evolution, but on the ``down'' path, reaching the lowermost
atom in the second node.  Now the third qubit of Q is sent.  It will
follow a path along the first two bifurcations that is determined by
the values of the first two qubits, and it will be absorbed in the
$|zero\rangle$ or $|one\rangle$ level of one of the atoms of the third
node, thus determining the path that the fourth qubit will follow, and
so on. After all the qubits of the Q register have been stored in the
respective nodes, a bus qubit initially in the state $|0\rangle$ is
sent through the apparatus.  Thanks to the presence of the other
qubits stored in the respective atoms, it will be directed to the
memory cell that is addressed by the Q register, where the cell's
value will be copied onto it. Now this bus qubit can follow the binary
tree backwards, exiting with the memory cell's value. To conclude the
protocol, all the atoms are made to emit their stored qubits backwards
with sequenced Raman transitions, starting from the last node and
progressing to the first.  The end result is that the memory cell's
content has been stored on the bus qubit, while all the qubits of the
Q register have been re-emitted and are again available.

\begin{figure}[hbt]
\begin{center}
  \epsfxsize=1.\hsize\leavevmode\epsffile{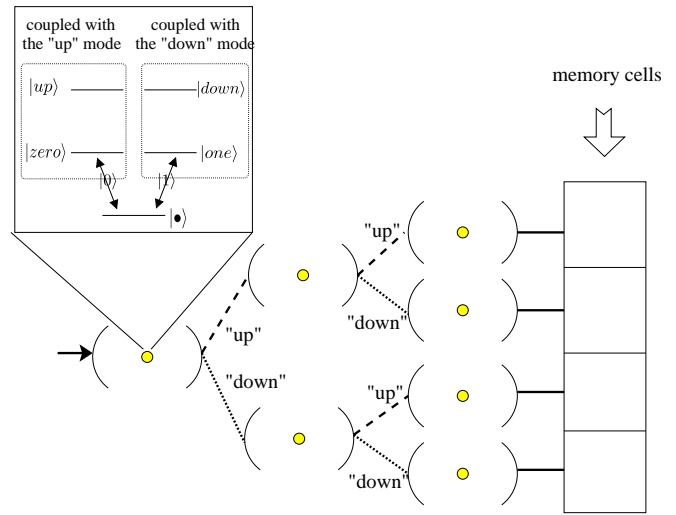}
\end{center}
\vspace{-.5cm}
\caption{Bucket-brigade implementation. At each node of a bifurcation
  tree, the three-state switch (the qutrit) consists of an atom in a
  cavity. Its level structure is given in the inset: The $|0\rangle$
  and $|1\rangle$ transitions couple only to incoming qubits which are
  in the $|0\rangle$ and $|1\rangle$ states respectively. When the
  first qubit arrives, depending on its state, it will be stored in
  the $|zero\rangle$ or $|one\rangle$ level. When a subsequent qubit arrives, it
  will be absorbed and re-emitted by the $|up\rangle$ levels or by the
  $|down\rangle$ levels depending on the earlier qubit's state. The $|up\rangle$
  levels are coupled only to the dashed arms of the binary tree,
  whereas the $|down\rangle$ levels are coupled to the dotted arms, so that
  all the subsequent qubits are routed depending on the state of the
  previous qubits. The extra atomic levels needed to perform Raman transitions
are not shown in the picture.} \labell{f:bb11}
\end{figure}

Because the bucket-brigade qRAM operates by sequential coupling of
qutrits, it takes $O(n^2)$ steps to retrieve one of $2^{n}$ memories
coherently.  In the discussed implementation the nodes of the binary
tree are atoms coupled via photons through Raman pulses. Alternatively
they could be solid-state artificial atoms such as superconducting
qubits or electron spin quantum dots, for which a variety of tunable
coupling schemes that allow the desired interactions have been
designed \cite{coupling1, coupling2, coupling3}.  All such schemes
will introduce errors, both via inaccuracies in the application of the
classical fields required to induce interactions between qutrits, and
via interaction with the environment.  A key requirement is that the
probability of passive states $|\bullet\rangle$ being inadvertently
excited to active states $|{zero} \rangle,|{one}\rangle$ be small
enough that signals are not routed along the wrong path.  Essentially,
as long as the error rate is significantly less than one over the
number of steps required in a memory call (i.e. $O(n^2)$) then the
coherent memory call goes through with high probability.

Another way to think of the resilience of the qRAM in the
face of noise and error is the following.
If the memory address register is initially in a superposition
of a large number, e.g., all, of the memory sites, then all
pieces of the qRAM circuit will be used during the coherent
memory call.  Because the bucket-brigade scheme for calling
a single memory involves only ${n}$ qutrits, however,
{\it in each component of the superposition} only ${n}$ qutrits
will be active.    Such superpositions are typically highly
robust in the face of noise and loss \cite{ensemble1, ensemble2}.

\section{Conclusions}\labell{s:concl}
Random access memory forms an integral part of computers and data
processing protocols, whether classical or quantum.  This paper showed
how quantum random access memories (qRAMs) might be implemented by
using quantum optical and solid-state quantum information processing
techniques.  We proposed a new paradigm for constructing both
classical and quantum RAMs, the bucket-brigade paradigm, in which the
degree of fanout required to perform a memory call can be reduced
exponentially, from $O(2^n)$ to $O(n)$.  As a result, the
memory-addressing portion of classical RAMs might be made more energy
efficient, and large quantum RAMs might be constructed without
recourse to expensive and difficult quantum error correction
techniques.  Such large-scale qRAMs could prove useful for fast
pattern recognition algorithms and for communication and computation
protocols in which one party wishes to access and process information
without the provider of the information knowing who accessed the
information or even what the information was.

\acknowledgements We thank Andrew Childs for pointing out
references~\cite{algorithm1,algorithm2,algorithm3,algorithm4}.  VG
acknowledges finantial support of Centro Ennio De Giorgi of Scuola
Normale Superiore of Pisa.


\begin{references}
\bibitem{nota}
Depending on the practical limitations that might be present in an
actual implementation, it is also possible to modify the above scheme
in order to remove any dependence on an external register A.  In fact,
instead of employing $2^{n}$ spatial modes in the bus photon, we can
use twice as many and encode the first qubit of A directly on the
bus photon's degree of freedom (using dual-rail logic). This means
that, once we have obtained the unary encoding of the bus photon, we can
transfer the first qubit of the memory into a dual-rail encoding on
the $2^{n}$ modes. We thus have to deal with $2\times 2^{n}$ modes, but we
can reverse the algorithm of Fig.~\ref{f:polarizz} separately on each
of these two groups. We end up with a single photon in two modes,
which encodes the value of qubit associated with the address
$|\Psi\rangle_Q$.
\bibitem{NOTAdue}
In fact, the
Heisenberg evolution of a cavity is given by~\cite{loudon}
\begin{eqnarray}
b_l&=&\kappa(\varphi)\; a_l+e^{i\varphi}\sigma(\varphi)\; a_r\;,\nonumber 
\\ b_r&=&\sigma(\varphi)\; a_l+\kappa(\varphi) \;a_r
\labell{cavity}\;,
\end{eqnarray}
where $a_l$ and $a_r$ are the cavity input fields (from the left and
from the right respectively), $b_l$ and $b_r$ are the cavity output
fields (from the left and from the right), and where $\kappa$ and
$\sigma$ are the cavity transmissivity and reflectivity. These
quantities are given by
$\kappa(\varphi)=\frac{\sqrt{1-\tau}(e^{i\varphi}-1)}{1-e^{i\varphi}(1-\tau)}$,
$\sigma(\varphi)=\frac{\tau}{1-e^{i\varphi}(1-\tau)}$
where $\tau$ is the transmissivity of the two beam splitters (left and
right) coupling the cavity to the external modes, and $\varphi$ is its
internal phase shift. The probability that a photon crosses the cavity
from left to right is given by $p=\langle {b_r}^\dag b_r\rangle$,
where initially only the mode $a_l$ is populated (with a single
photon). Using Eq.~(\ref{cavity}), we find
\begin{eqnarray}
  p=|\sigma(\varphi)|^2=\frac{\tau^2}{(1-\tau)^2-2(1-\tau)\cos(\varphi)+1}
  \labell{p}\;.
\end{eqnarray}
By simply decreasing $\tau$, the transmissivity peaks of Eq.~(\ref{p})
can be narrowed at will, thus increasing the phase selectivity of the
cavity. The width of the peaks can be calculated by imposing
$|\sigma(\varphi)|^2=1/2$, which has solution
$\varphi=\arccos\left(1-\frac{\tau^2}{2(1-\tau)}\right)\simeq\tau$.
Hence, in order to get a phase selectivity of the order of
$\pi/2^{n}$, we need an exponentially small transmissivity $\tau\simeq
2^{-n}$ in the cavity's beam splitters.
\end{references}
\end{document}